\documentclass[final,1p,times]{elsarticle} 

\usepackage{graphicx}
\usepackage{amssymb} 
\usepackage{amsthm} 
\usepackage{lineno}

\newcommand{\pt}{$p_{\rm T}$}
\newcommand{\Raa}{$R_{\rm AA}$}

\journal{Nuclear Physics A} 

\begin{document}

\begin{frontmatter} 

\title{Neutral meson production in $pp$ and $Pb-Pb$ collisions at the LHC
measured with ALICE}

\author{D.Peresunko (for the ALICE\fnref{col1} Collaboration)}
\fntext[col1] {A list of members of the ALICE Collaboration and acknowledgements
can be found at the end of this issue.}
\address{RRC ``Kurchatov institute, Kurchatov sq.,1, Moscow}


\begin{abstract} 
We present spectra of $\pi^{0}$, $\eta$ and $\omega$ mesons 
in pp collisions and $\pi^{0}$ mesons in Pb-Pb collisions measured with ALICE at LHC energies.
The $\pi^{0}$ and $\eta$ mesons are reconstructed via their two-photon decays by two complementary methods, 
using the electromagnetic calorimeters and photon conversion technique; both measurements show perfect agreement. 
We measure the nuclear modification factor ($R_{AA}$) of $\pi^{0}$ production in Pb-Pb collisions at
different collision centralities and compare with lower energy results and theoretical predictions. 
\end{abstract} 

\end{frontmatter} 


\section{Introduction}

Neutral mesons, in particular $\pi^0$ and $\eta$ mesons can be
reconstructed and clearly identified using two-photon decays in a very wide \pt
~range. This makes them an excellent tool both for testing QCD predictions 
of meson yields in pp collisions and estimating parton energy loss in Pb-Pb collisions.
With these measurements at LHC energies one can access the Parton Distribution Functions 
(PDF) and Fragmentation Functions (FF) in new kinematic regions and provide 
further constraints on their low-$x$ (low-$z$) parts.
Meson production at LHC 
energies is dominated by gluon fragmentation up to transverse momentum $p_T \sim 100$ GeV/$c$ \cite{gluonFF}, 
a kinematic regime which makes them ideal to constrain quark as well as the lesser know gluon FF. 
Measurement of the nuclear modification factor ($R_{AA}$) of $\pi^{0}$ production 
in Pb-Pb collisions provides a clear pattern for identified particles and is free from 
ambiguity, such as that found in the separation of protons, kaons and pions in charged hadron suppression.

\section{ALICE setup}


Photons from neutral meson decays can be detected either in the electromagnetic calorimeters 
PHOS  or  EMCAL, or can convert on the material of the inner detectors
and be reconstructed via $e^+e^-$ pair in the central tracking system. 
All three identification methods can be used to provide three independent measurements simultaneously. 
Below results obtained with PHOS and photon conversion method are presented.

The electromagnetic calorimeter PHOS \cite{PHOS} is made of $PbWO_4$ crystals and has fine 
granularity (crystal size $2.2\times 2.2\times 18$ cm$^3$ installed at the distance 4.6 m from
the interaction point), good energy and  position resolutions. PHOS covers $|\eta|< 0.125$ in 
pseudorapidity and $60^\circ$ in azimuthal angle with the 3 modules currently installed. Effort 
was made to reduce the material budget in front of PHOS as much as possible: $0.2\cdot X_0$. 
Thanks to these conditions PHOS can resolve 
two photons from $\pi^0$ decay up to $p_{\rm T}^\pi \sim 50$ GeV/$c$.

In the case of conversion technique, tracks are reconstructed in the ALICE tracking system, 
consisting of the Inner Tracking System (ITS) \cite{ITS} and the Time Projection Chamber (TPC).
The candidate track pairs are selected using a secondary vertex (V0) finding algorithm. 
Further refining of the photon sample and rejection of contaminations like $K_S^0$, $\Lambda$, $\bar\Lambda$ 
is performed using electron identification via $dE/dx$ and a cut on the opening angle of the pairs.

\section{pp collisions}

\begin{figure}[t]
\begin{center}
\includegraphics[width=0.45\textwidth]{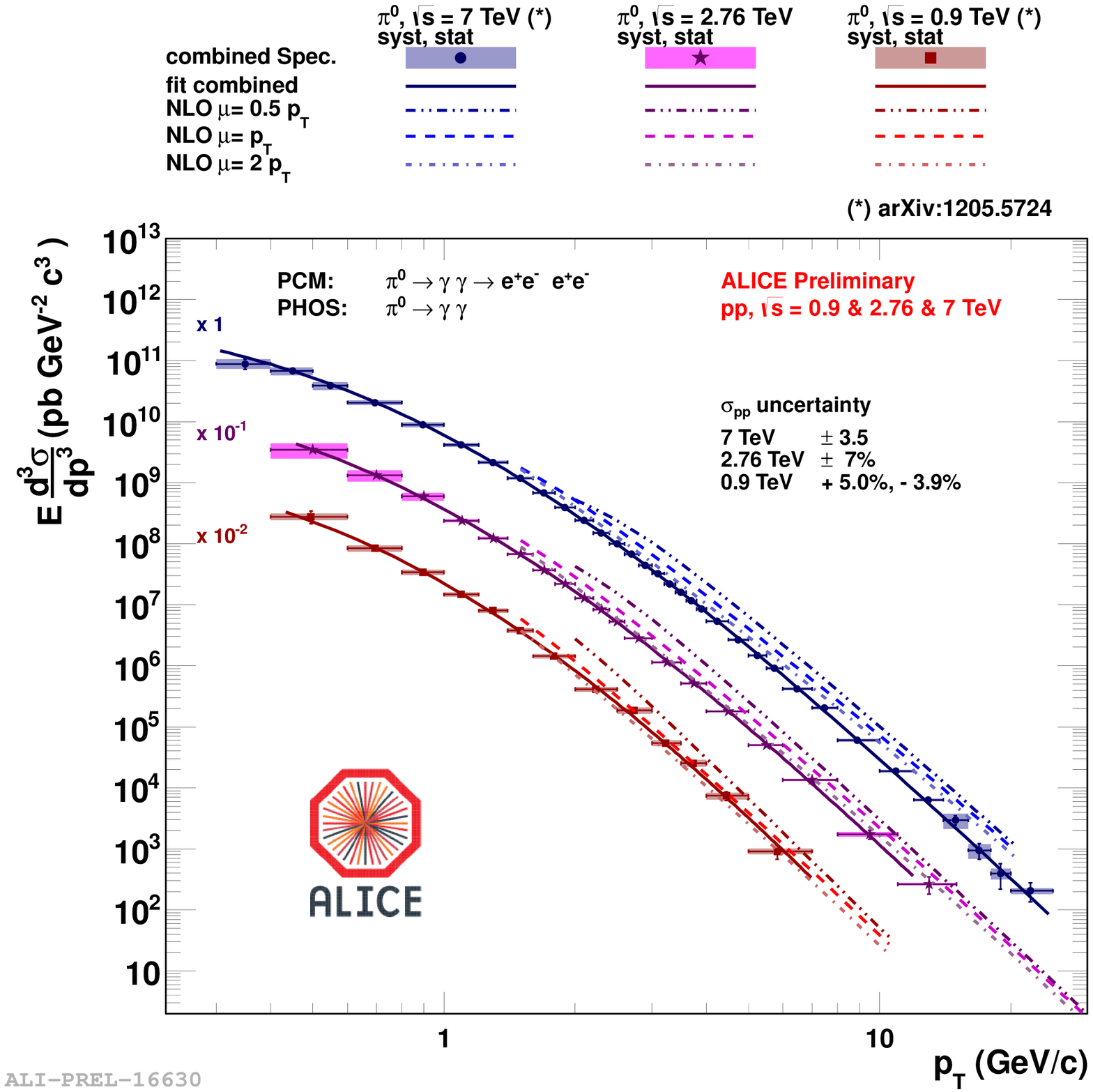}
\hfill
\includegraphics[width=0.45\textwidth]{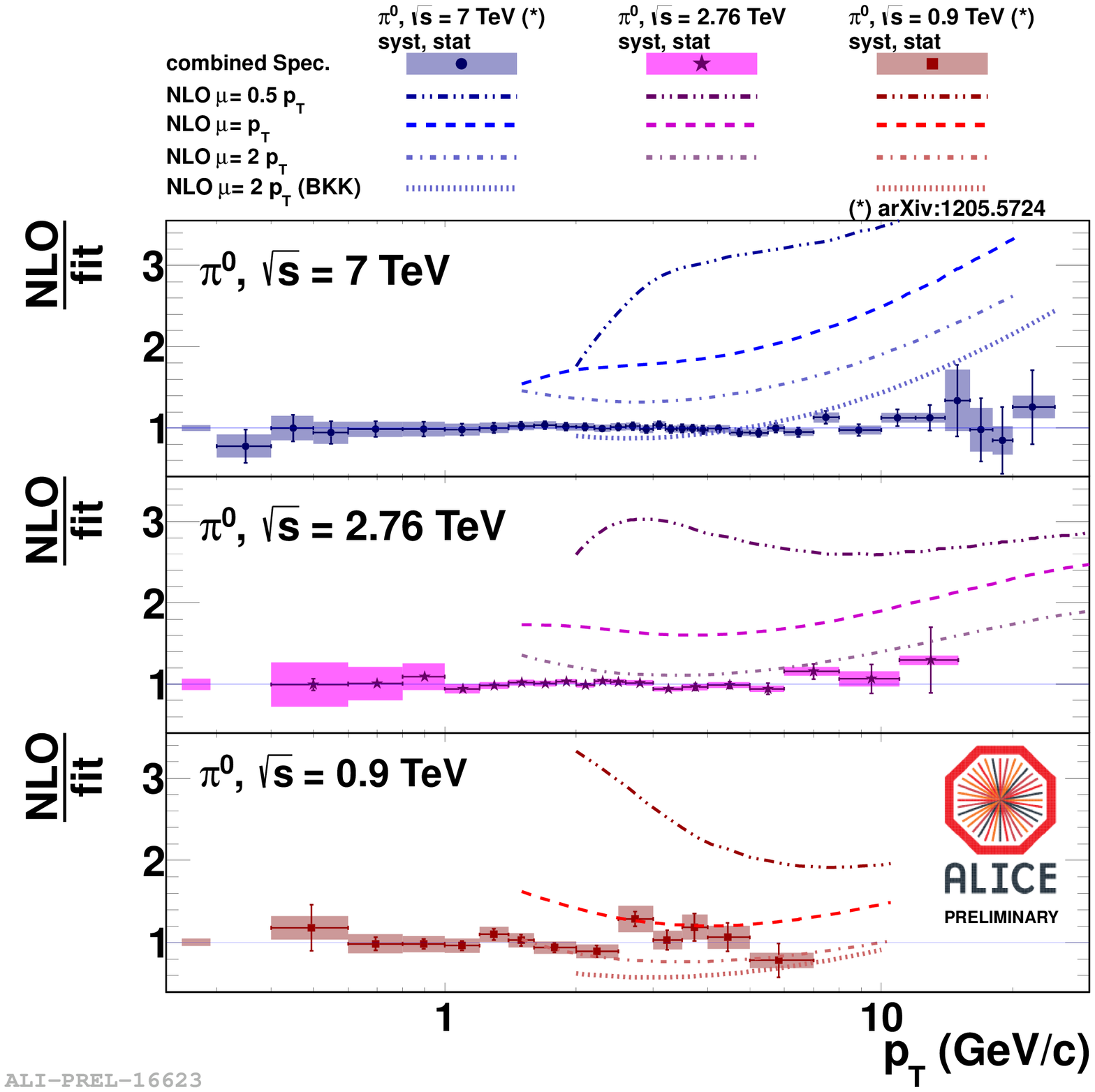}
\end{center}
\caption{(Color online). Left plot: Spectra of $\pi^{0}$ measured in pp collisions at 3 colliding energies, 
compared to the NLO pQCD predictions and Tsallis fit. Right plot: Ratio data/Tsallis fit compared to the
ratio NLO pQCD predictions/Tsallis fit.  Theory curves reproduce data at $\sqrt{s}=0.9$ TeV but overpredict
at 2.76 and 7 TeV \cite{pp-pi0}. }
\label{fig:pi0pp}
\end{figure}

ALICE has measured with the 2010 data set $\pi^0$ spectra at three colliding energies, see fig.\ \ref{fig:pi0pp}. 
Collected statistics corresponds approximately to an integrated luminosity of
0.14 nb$^{-1}$, 0.7 nb$^{-1}$ and 5.6 nb$^{-1}$ 
for energies $\sqrt{s}=0.9$, 2.76 and 7 TeV, respectively. Spectra were measured independently 
with PCM and PHOS, results which were in agreement and were subsequently combined into a final spectrum. 
The measured spectra were compared with NLO pQCD predictions \cite{pi0-theory}. For the better
comparison we fit spectra with the Tsallis parameterization (solid lines in the left plot of fig.\ \ref{fig:pi0pp}) 
and present the data-to-fit ratio with the predictions-to-fit ratio in the right plot. We find that theoretical calculations  
reproduce $\pi^0$ spectrum at $\sqrt{s}=900$ GeV but considerably overestimate the 
$\pi^{0}$ yield at higher energies.

Using the same data, ALICE measured the spectrum of $\eta$ meson in pp collisions at the same collision 
energies reported above \cite{pp-pi0}.
Comparison of the $\eta$ meson spectra with theoretical predictions is similar to that of $\pi^0$:
 NLO pQCD predictions reproduce 
data at $\sqrt{s}=900$ GeV but overestimate yields at higher energies. From the measured spectra, 
the ratios $\eta/\pi^0$ were constructed. 
These ratios measured at the three LHC energies agree with each other and with measurements at lower energies. 
In contrast to the spectra, the meson ratios agree well with NLO predictions.

\begin{figure}[t]
\begin{center}
\includegraphics[width=0.44\textwidth]{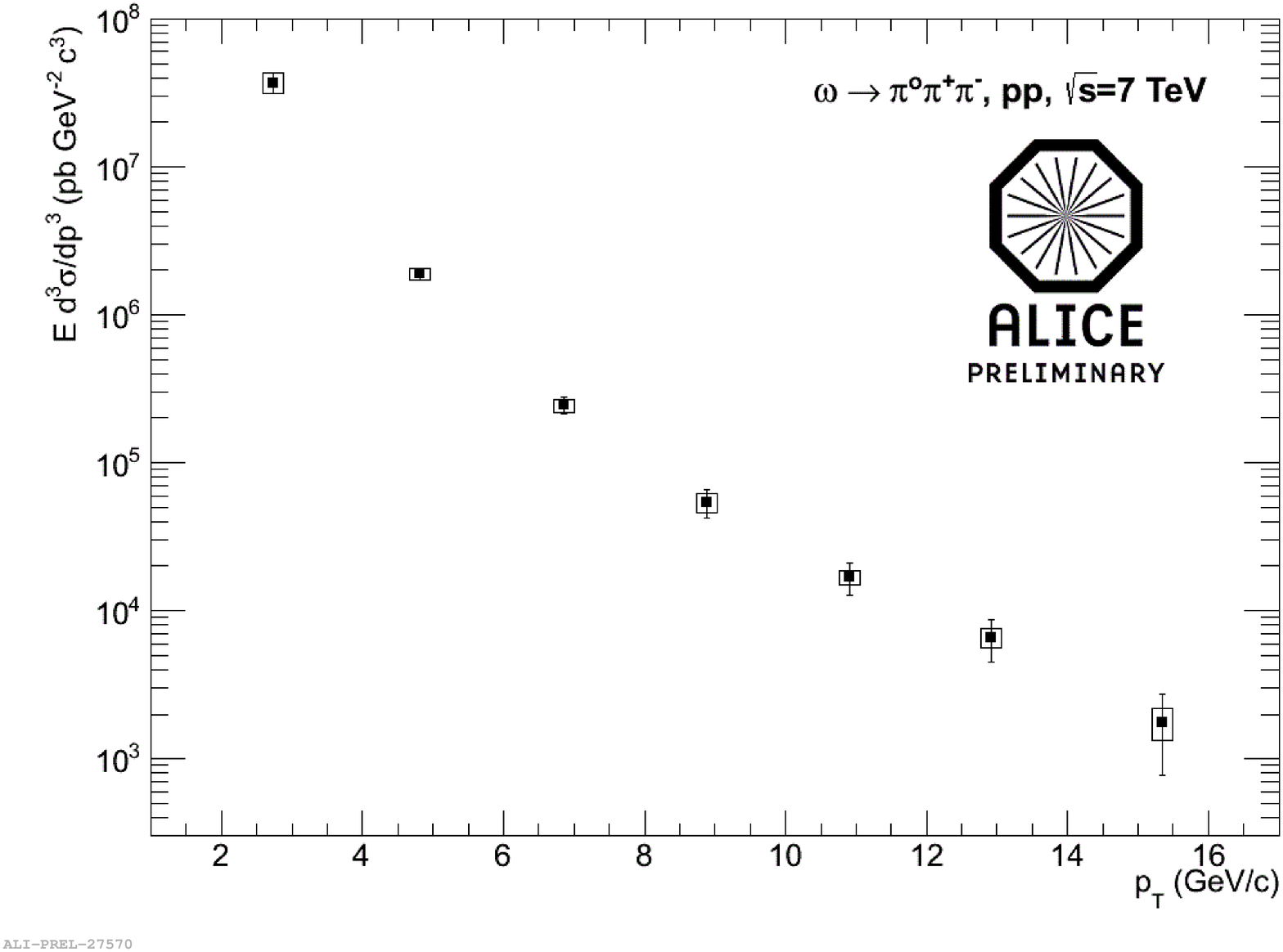}
\hfill
\includegraphics[width=0.44\textwidth]{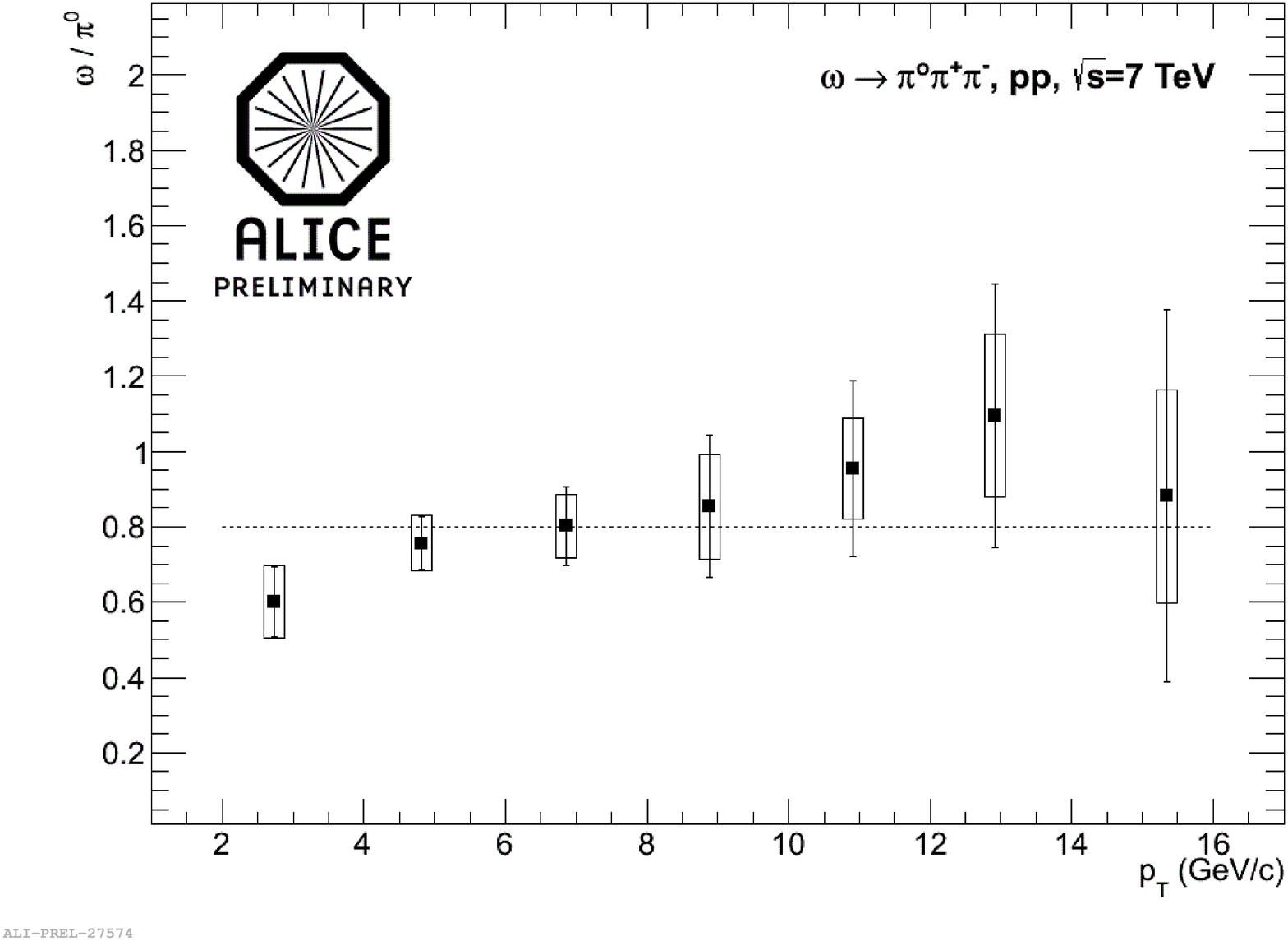}
\end{center}
\caption{Left: spectrum of $\omega$ meson measured in pp collisions at $\sqrt{s}=7$ TeV. 
Right: Ratio $\omega/\pi^0$ measured in pp collisions at $\sqrt{s}=7$ TeV. Dashed line -- 
world average of asymptotic ratios measured in pp collisions at lower energies.}
\label{fig:omega}
\end{figure}

In addition to $\pi^0$ and $\eta$ meson spectra, the spectrum of $\omega$ meson was measured 
in pp collisions at $\sqrt{s}=7$ TeV. In the low multiplicity environment of pp collision the
best decay channel is $\omega \to \pi^0\pi^+\pi^-$ with the branching ratio of 89.2\%. We present 
the results obtained with the 2010 data set. 400 Mevents were analyzed corresponding approximately to 
an integrated luminosity of 6 nb$^{-1}$. Charged pions were reconstructed in the central tracking system 
and $\pi^0$s were detected in PHOS. The spectrum is shown in fig.\ \ref{fig:omega}, left plot.
Unfortunately, as the $\omega$ meson fragmentation function is not yet known currently, there are 
no NLO predictions available. The measurement of the $\omega$ spectrum in a wide \pt 
~range by ALICE can be used as an input for theoretical constrains on parametrization of the $\omega$ FF. 
The ratio 
$\omega$/$\pi^{0}$ is shown in the right plot. We found that slopes of the spectra are close at high \pt$\gtrsim 4$ GeV/$c$, 
and the relative yield of $\omega$ agrees with world average calculated for pp collisions at lower energies 
$\sim 0.8$ \cite{omega-PHENIX} shown with dashed line.

\section{Pb-Pb collisions}

\begin{figure}[t]
\begin{center}
\includegraphics[width=0.43\textwidth]{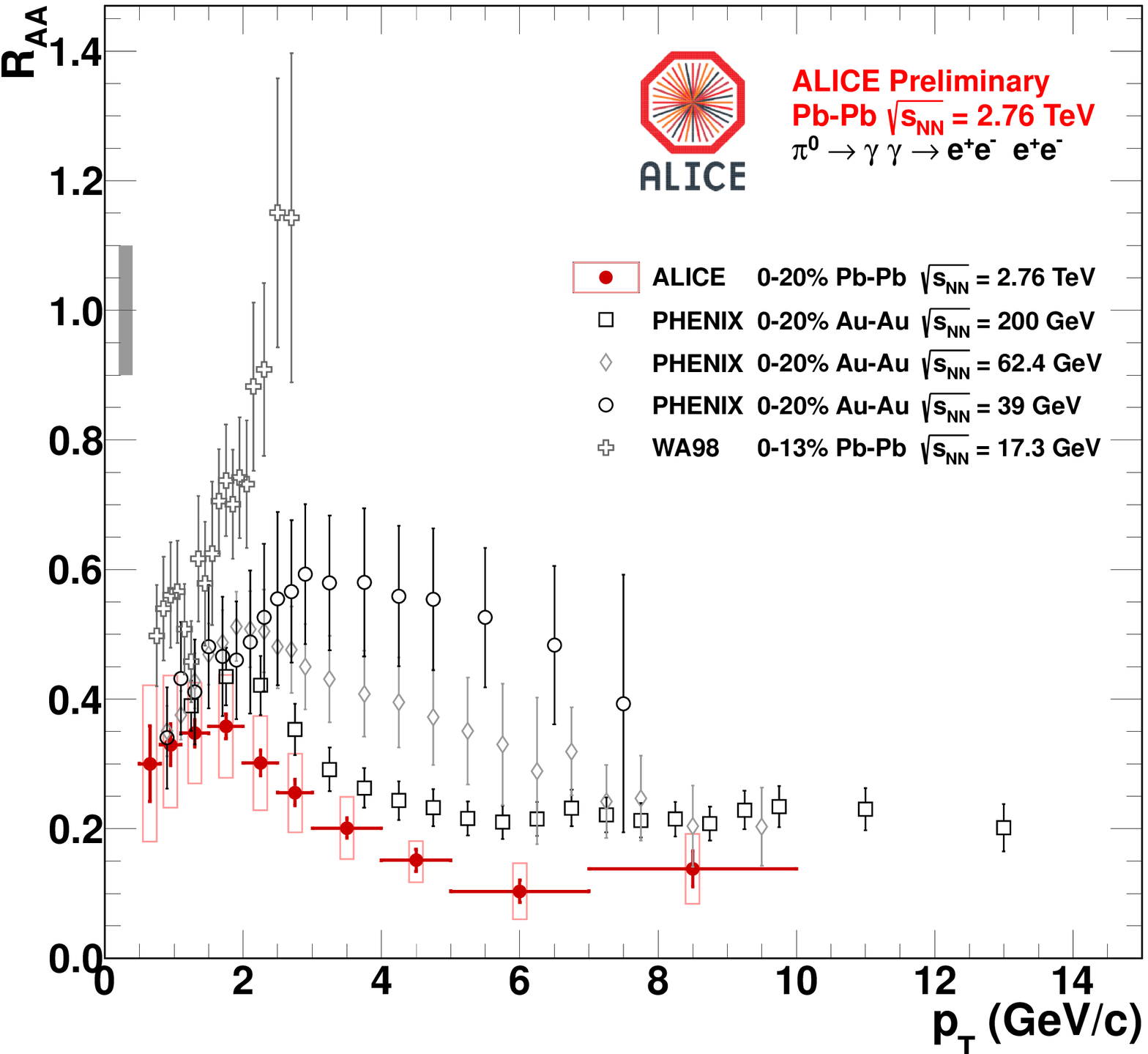}
\hfill
\includegraphics[width=0.43\textwidth]{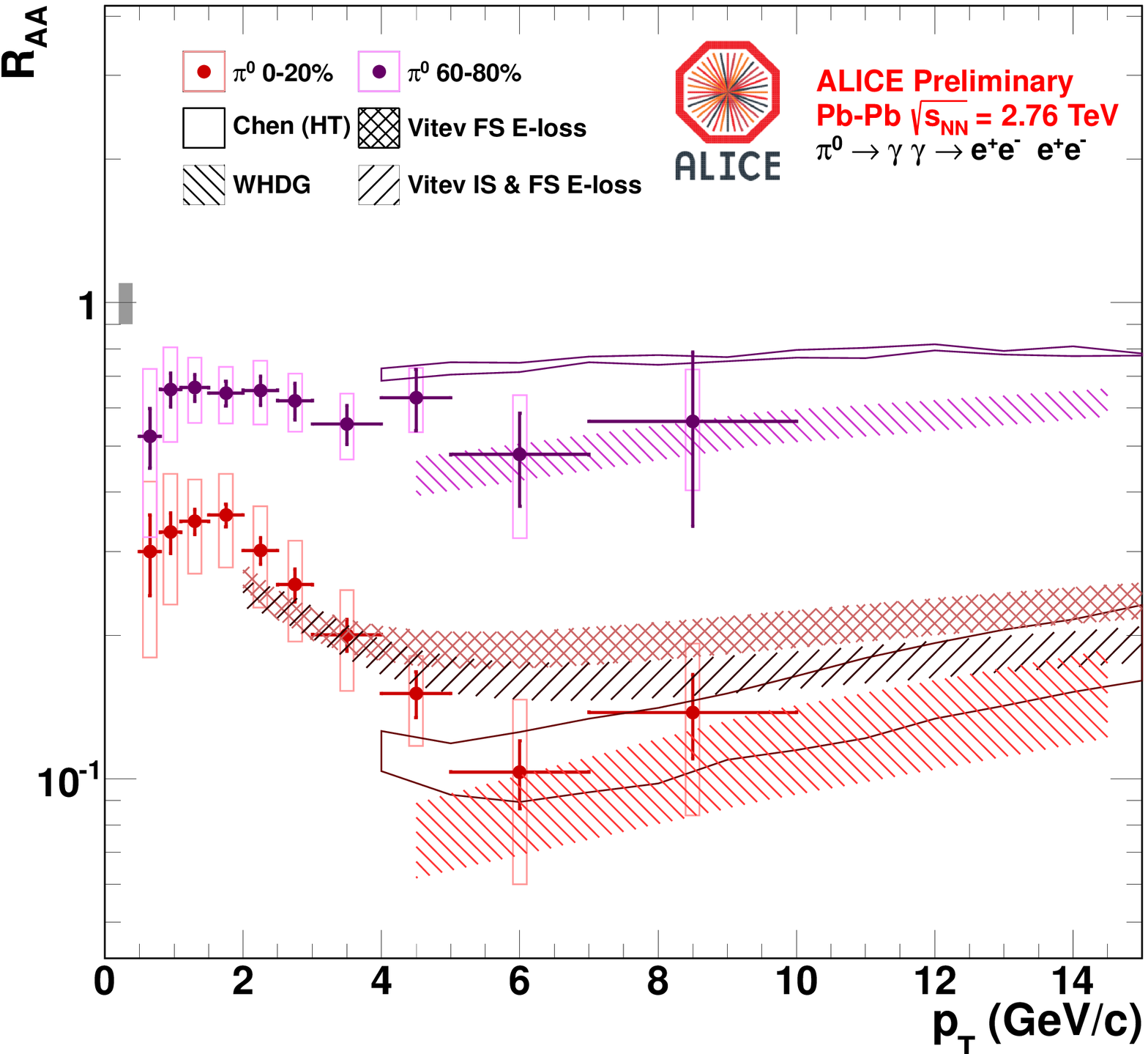}
\end{center}
\caption{(Color online) Left plot: $\pi^0$ nuclear modification factor (\Raa) measured in 20\% most central
Pb-Pb collisions compared to \Raa ~in Pb-Pb and Au-Au collisions at lower energies. Right plot: \Raa ~in 20\% 
of most central and 60-80\% peripheral collisions compared to several theoretical predictions.}
\label{fig:Raa-RHIC}
\end{figure}

A-A collisions are characterized by a high detector occupancy
resulting in considerable probability of cluster overlaps and large combinatorial 
background. Nevertheless ALICE extracted
$\pi^0$ spectra in several centrality bins in Pb-Pb collisions at $\sqrt{s_{NN}}=2.76$ TeV. 
Using baseline measurements in pp collisions at the same energy, a nuclear modification
factor \Raa~was constructed. We present \Raa ~measured in 20\% of the most central collisions (fig.\ \ref{fig:Raa-RHIC}, 
left plot). It is compared with \Raa ~measured at lower energies with WA98 \cite{WA98} and PHENIX \cite{PHENIX} 
collaborations. 
We find that starting from the lowest RHIC energy, \Raa ~has approximately the same shape. 
Furthermore, the amount of suppression gradually increases with increasing energy of the collision. 

We compare the measured nuclear modification factors with several theoretical predictions to be described hereafter 
(see fig.\ \ref{fig:Raa-RHIC}, right plot).
The WHDG model \cite{Horowitz:2007nq} takes into account collisional and radiative parton
energy loss and geometrical path length fluctuations. The color charge
density of the medium is assumed to be proportional to the number of
participating nucleons from the Glauber model. The WHDG model reproduces both strength of the 
suppression and its centrality dependence. 
Higher twist calculations \cite{Chen:2011vt} differ from the WHDG model in the
implementation of the medium properties. In addition, the
space-time evolution of the medium is described with a $3+1$
dimensional ideal hydrodynamics. Its predictions agree with
data in central collisions but fail to reproduce the centrality
dependence. 
Two calculations \cite{Sharma:2009hn} consider two cases: 
incorporate only final-state parton energy loss; take into account
energy loss of the incoming parton and the broadening of the transverse momenta of the
incoming partons in the cold nuclear matter.
Both calculations agree with data, however a somewhat improved agreement is reached
 if both initial state and final state
emissions are taken into account.

\section{Summary}

We presented  neutral meson spectra measured by ALICE collaboration in pp collisions at three collision energies and \Raa~of $\pi^0$
measured in Pb-Pb collisions. Spectra measured with two different techniques, with PHOS and
with photon conversion method, show perfect agreement. We found that NLO pQCD predictions describe well $\pi^{0}$ and $\eta$  
production in pp at $\sqrt{s}=0.9$ TeV, but overestimate their production at $\sqrt{s}=2.76$ and 7 TeV. 
We presented the spectrum of $\omega$ meson in pp collisions at $\sqrt{s}=7$ TeV and found $\omega$/$\pi^0$ ratio in agreement
with ratios measured at lower energies. Suppression of $\pi^0$ in Pb-Pb collisions at $\sqrt{s_{NN}}=2.76$ TeV is stronger than
the one measured at RHIC energies.

This work was partially supported by Russian RFBR grant 12-02-91527.

\end{document}